\documentclass[submission,copyright,creativecommons]{eptcs}
\providecommand{\event}{WACA 2025} 
\providecommand{\thisvolume}[1]{this volume of EPTCS, Open Publishing Association}
\usepackage{iftex}
\usepackage{todonotes}
\usepackage{algorithm2e}
\usepackage{subcaption}
\usepackage{multirow}

\hypersetup{
colorlinks=true,
linkcolor=black,
citecolor=black,
anchorcolor=black,
filecolor=black,
urlcolor=black
}

\usepackage{comment}

\providecommand{\thisvolume}[1]{this volume of EPTCS, Open Publishing Association}

\makeatletter
\usepackage{fancyvrb}
\usepackage{amssymb}
\newcommand{\ygg@basicalert}[2]{\fbox{\bfseries\sffamily\scriptsize#1}{\sf\small$\blacktriangleright$\textit{#2}$\blacktriangleleft$}}
\newcommand{\annote}[2]{\ygg@basicalert{\textcolor{red}{\textsc{#1}}}{\textcolor{red}{#2}}}

\newcommand{\cesar}[1]{\ygg@basicalert{\textcolor{blue}{\textsc{César}}}{\textcolor{blue}{#1}}}

\makeatother

\newcommand\DM[1]{\textcolor{black}{#1}}

\newcommand{\henrique}[1]{\textcolor{black}{#1}}

\newcommand{\ie}{\textit{i.e.,}\xspace}
\newcommand{\eg}{\textit{e.g.,}\xspace}
\newcommand{\etal}{\textit{et al.}~}
\newcommand{\wrt}{\textit{w.r.t.}~}

\newcommand{\app}{\textsf{TeaStore} application\xspace}
\newcommand{\adapApp}{\textsf{Adaptable TeaStore}\xspace}
\newcommand{\tool}{\textsf{EnCoMSAS}\xspace}

\ifpdf
  \usepackage{underscore}         
  \usepackage[T1]{fontenc}        
\else
  \usepackage{breakurl}           
\fi

\title{Adaptable Teastore with Energy Consumption Awareness: A Case Study}
\author{Henrique De Medeiros\footnote{Corresponding author: \email{henrique.de\_medeiros@telecom-sudparis.eu}} (1), Denisse Mu\~nante (2), Sophie Chabridon (1)\\ César Perdigão Batista (1), Denis Conan (1)
\institute{(1) SAMOVAR, Télécom SudParis, Institut Polytechnique de Paris, 
91120 Palaiseau, France\\
(2) ENSIIE \& SAMOVAR, \'Evry, France}
}
\def\titlerunning{Adaptable Teastore with Energy Consumption Awareness: A Case Study}
\def\authorrunning{H. De Medeiros et al.}
\begin{document}
\maketitle

\begin{abstract}
\textbf{[Context and Motivation]}
Global energy consumption has been steadily increasing in recent years, with data centers emerging as major contributors. This growth is largely driven by the widespread migration of applications to the Cloud, alongside a rising number of users consuming digital content. 
\DM{Dynamic adaptation (or self-adaptive) approaches appear as a way to reduce, at runtime and under certain constraints, the energy consumption of software applications.}

\noindent \textbf{[Question/Problem]}
\DM{Despite efforts to make energy-efficiency a primary goal in the dynamic adaptation of software applications, there is still a gap in understanding how to equip these self-adaptive software systems (SAS), which are dynamically adapted at runtime,  with effective energy consumption monitoring tools that enable energy-awareness. Furthermore, the extent to which such an energy consumption monitoring tool impacts the overall energy consumption of the SAS ecosystem has not yet been thoroughly explored.}

\noindent \textbf{[Methodology]}
To address this gap, \DM{we introduce the \tool (\textbf{En}ergy \textbf{Co}nsumption \textbf{M}onitoring for \textbf{S}elf-\textbf{A}daptive \textbf{S}ystems) tool that allows to gather the energy consumed by distributed software applications deployed, for instance, in the Cloud.
\tool enables the evaluation of energy consumption of SAS variants at runtime. It allows to integrate energy-efficiency as a main goal in the analysis and execution of new adaptation plans for the SAS.}

\noindent \DM{In order to evaluate the effectiveness of \tool and investigate its impact on the overall energy consumption of the SAS ecosystem, we conduct an empirical study by using the \textit{\adapApp} case study. \adapApp is a self-adaptive extension of the \app, a microservice benchmarking application. For this study, we focus on the recommender service of \adapApp. 
Regarding the experiments, we first equip \adapApp with \tool. Next, we execute \adapApp by varying workload conditions that simulate users' interactions. Finally, we use \tool for gathering and assessing the energy consumption of the recommender algorithms of \adapApp. 
To run these experiments, we use nodes of the Grid5000 testbed.}

\noindent \textbf{[Results]}
\DM{The results show that \tool is
\henrique{effective in collecting energy consumption of software applications}
for enabling dynamic adaptation at runtime. The observed correlation between CPU usage and 
energy consumption collected by \tool provides evidence supporting the validity of the collected energy measurements. Moreover, we point out, through \tool, }
that energy consumption is influenced not only by the 
\henrique{algorithmic complexity}
but also by the characteristics of the deployment environment. \DM{Finally, the results show that the impact of \tool on the overall energy consumption of the SAS ecosystem is comparatively modest with respect to the entire set of the \app's microservices.}
\end{abstract}

\section{Introduction}
Global energy consumption has been steadily increasing, with data centers emerging as major contributors. This growth is largely driven by the widespread migration of applications and businesses to the Cloud, alongside a rising number of users consuming digital content. To understand the connection between application behaviour and energy use, two key concepts are essential: energy-efficiency and energy-awareness.
Energy-efficient applications are designed to operate with minimal energy impact, while energy-aware applications are 
\henrique{designed} with a conscious understanding of their energy footprint.

One of the challenges in building energy-efficient applications lies in accessing their energy consumption data. In Cloud environments, where infrastructure is managed by providers and resources are delivered on a pay-per-use basis, developers often rely on metrics like CPU and memory usage to gauge performance. In this paper, we work at the software application level to compute the power consumption. 


This paper thoroughly analyses a self-adaptive application that integrates an energy consumption profiling framework. Specifically, we examine the energy consumption of different algorithms across heterogeneous CPU architectures.
We rely on the case study of the \adapApp~\cite{BDGLZZ25} application and consider different recommendation algorithms. This \henrique{application} 
simulates an online tea shop and supports architectural adaptations, making it an ideal candidate for studying energy-aware design in practice. We have extended the 
application with energy awareness. While, in the base application, the CPU load is monitored to determine resource consumption, we monitor the energy consumption of the application with an energy profiling tool.
Initially, we introduce the framework responsible for monitoring application-level energy consumption and then describe the experimental protocol based on \adapApp. Using this setup, we evaluate the energy consumption of multiple recommender algorithms against their algorithmic complexity.
In a second phase, we extend the protocol to assess the adaptive capabilities of the recommender microservice with respect to energy consumption. In this context, we identify the most energy-efficient recommender algorithm and evaluate the trade-off between energy consumption and performance by dynamically switching at runtime between an energy-efficient algorithm and a performance-oriented one. Our analysis considers three configurations: the adaptive version, the non-adaptive version, and the original TeaStore implementation. Finally, we assess the overall impact of our approach on system-wide energy consumption.

This paper is organised as follows: Section~\ref{sec:background} provides background information and defines the key concepts used throughout the study. Section~\ref{sec:experiment} describes the proposed extension of the \adapApp application. Section~\ref{sec:experimentv2} specifies research questions and the designed experiments to address these research questions. Section~\ref{sec:results} presents and analyses the results obtained. Section~\ref{sec:threats} discuss about the threats to validity. Section~\ref{sec:relatedwork} reviews recent related works in the field. Finally, Section~\ref{sec:conclusion} concludes the paper.
\label{sec:introduction}

\section{Background}\label{sec:background}
This section introduces the tools used for monitoring the energy consumed by software applications. Next, we introduce the \app. Finally, we explain the \adapApp application which is the extension of the \app to support its dynamic adaptation at run-time.

\subsection{Energy Monitoring Tools}

The energy consumption of software applications can be measured through various approaches. Jay~\etal~\cite{2023Jay} categorise these approaches into four main groups: external devices, intra-node devices, hardware sensors with software interfaces and power/energy software models.



In Cloud computing environments, the hardware-based methods often fall short, measuring the total energy consumption of a machine or a computing node without offering sufficient granularity to attribute energy consumption to individual applications or processes running within the shared infrastructure.
To address this limitation, software power meters have emerged as a practical alternative. These tools do not require external hardware and estimate energy consumption through mathematical models. Most of them rely on existing hardware sensors and software interfaces, such as Intel’s RAPL (Running Average Power Limit) for CPUs and Nvidia’s NVML (NVIDIA Management Library) for GPUs~\cite{cadorel2024protocolpower}.

Based on the previous interfaces and memory consumption indicators, other software tools estimate power consumption at the process level. Notable examples of software power meters include Linux perf, Kepler\footnote{\url{https://sustainable-computing.io/}} (targeted for Kubernetes environments), PyJoules\footnote{\url{https://github.com/powerapi-ng/pyJoules}}, PowerAPI\footnote{\url{https://powerapi.org/}}, and Scaphandre\footnote{\url{https://github.com/hubblo-org/scaphandre}}. PyJoules, PowerAPI, and Scaphandre are particularly interesting for their ability to measure power consumption 
at the level of detail required for improving the energy-awareness of software applications.

\subsection{The TeaStore Application}

\begin{figure}[h]
    \centering
    \includegraphics[width=.6\linewidth]{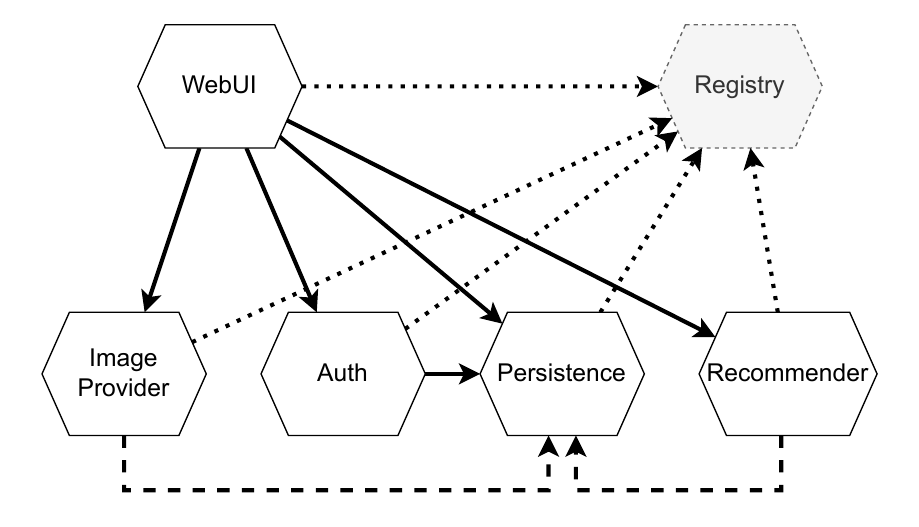}
    \caption{TeaStore Architecture, adapted from~\cite{2018TeaStore}}
    \label{fig:teastore}
\end{figure}

TeaStore~\cite{2018TeaStore} is a distributed microservice application that emulates an online store for buying tea and tea related utilities.
%
Figure~\ref{fig:teastore} depicts its architecture, composed of six microservices.

The \textsf{WebUI} service is the entrypoint for users' requests and serves the application web interface to support interactions. The \textsf{Authentication} service manages user login and session validation, thus ensuring secure access to the system. The \textsf{Recommender} service leverages rating algorithms to suggest products to users. Recommendations are generated based on the user behaviour, \eg the products being viewed by the user, the items in the shopping cart or the purchasing habits of other users. The \textsf{Persistence} service provides the information about the products and the user's purchases. The \textsf{Image provider} service is responsible for delivering the images used throughout the application. It dynamically adjusts image sizes based on the display context and incorporates caching mechanisms to ensure quick access to frequently requested images. The \textsf{Registry} service connects all active service instances by relying on their IP addresses or host-names and corresponding ports. It enables service discovery, thus allowing microservices to seamlessly locate and communicate with each other. All these services communicate using REST and Netflix Ribbon client-side load balancer.

\subsection{The Adaptable TeaStore Application}
\label{subsec:theadaptableteastore}

\begin{figure}[h]
    \centering
    \includegraphics[width=1\linewidth]{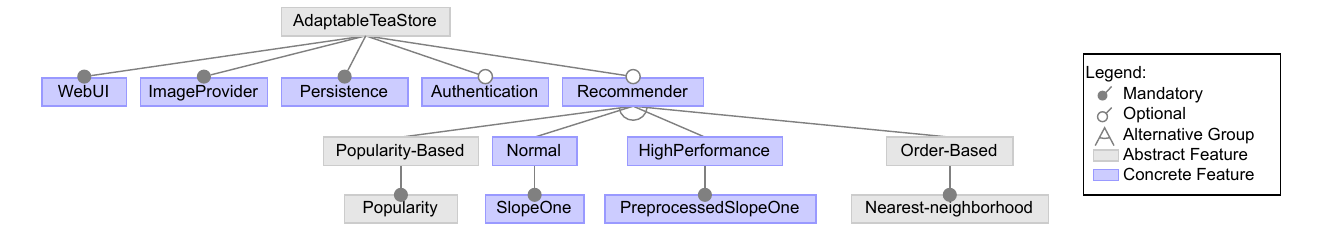}
    \caption{The feature model of \adapApp}
    \label{fig:fm}
\end{figure}

\adapApp~\cite{DGLZ25} is an extension of the \app, and is designed to support dynamic adaptations when some particular events are triggered. 
Mandatory and optional services are identified (see Figure~\ref{fig:fm}). \textsf{WebUI}, \textsf{Persistence}, and \textsf{Image Provider} correspond to the mandatory services, \ie they are required for all configurations of the \app, while \textsf{Recommender} and \textsf{Authentication} correspond to the optional services, \ie they could be present or not in a configuration of the \app. In addition, these optional services can also be adapted to adjust their resource consumption.  


The \textsf{Recommender} service implements four recommendation algorithms, namely \textit{Popularity}, \textit{Slope One}, \textit{Preprocessed Slope One}, and \textit{Order-Based} methods.
The algorithm is chosen at runtime, for instance, to alleviate/decrease the CPU usage.

After analysing the source code of \adapApp, we determine that only three potential configuration modes are possible. The \textit{Normal Mode}, or \textit{Default Recommender}, restores the default recommendation algorithm, which is the \textit{Slope One} algorithm. The \textit{High Performance Mode} configures the system to use the \textit{Pre-processed Slope One} algorithm, aiming for maximum performance. And, the \textit{Low Power Mode}  disables the recommendation algorithm, \ie the recommender is deactivated. 
According to the source code, the popularity-based and the order-based algorithms are not enabled for adapting the \adapApp. Therefore, they are shown in Figure~\ref{fig:fm} as abstract features that will not be included in the dynamic adaptations of the 
\adapApp. 


Mode transitions are triggered based on CPU usage: when CPU usage exceeds 50\%, the system switches to \textit{Low Power} mode. 
If usage drops below 50\%, it returns to \textit{Normal} mode, \ie the Recommender is switched-on with the activation of the \textit{Slope One} algorithm. 
Finally, in the current version of the \adapApp, the \textit{High Performance} mode is not activated for any event.


\section{Extending Adaptable Teastore with Energy Consumption Awareness}\label{sec:experiment}

\begin{figure}[!ht]
    \centering
    \includegraphics[width=1\linewidth]{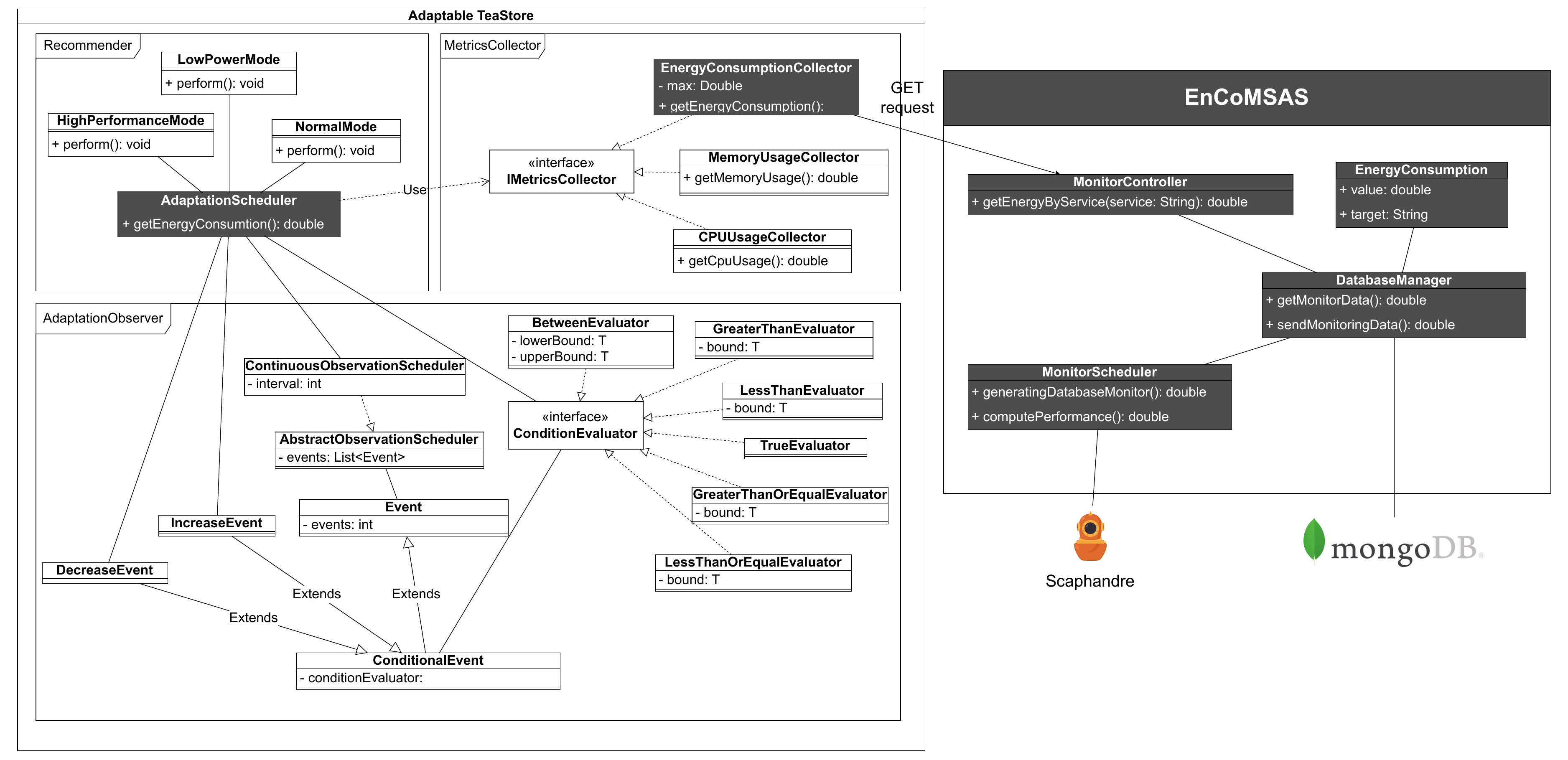}
    \caption{The extended architecture of \adapApp }
    \label{fig:architecture}
\end{figure}

To enable energy-awareness, we introduce the so-called \tool (\textsf{En}ergy \textsf{Co}nsumption \textsf{M}onitor for \textsf{S}elf-\textsf{A}daptive \textsf{S}ystems) tool,
represented as the grey component in the top-right corner of Figure~\ref{fig:architecture}.
It serves as a higher-layer energy consumption monitoring and analysis tool. It mainly allows to configure the monitoring frequency, enables the computation of the energy consumed by the target software application, and facilitates the analysis of the energy consumed by the
different variants of the software application.
The  \tool
tool is designed not only for a specific microservice/application, \eg \adapApp, but it is generic and can be used for any application.
For this,  \tool 
receives the name of the software application, it then 
collect different metrics, \eg power consumption, for this application, and finally computes energy consumption based on these metrics.

In our experiments,  \tool
uses \textsf{Scaphandre} for computing energy consumption, however  \tool
is designed for being easily extended by adding another monitoring tool. We conducted the same set of experiments using PowerAPI. However, given the similarity of the results reported by both tools~\cite{2023Jay}, we chose to proceed with Scaphandre for the remainder of our study.
\textsf{Scaphandre} reads instantaneous power consumption measurements of running applications, as well as of the entire machine, that are exposed 
with a Prometheus exporter. We store these energy values in a data storage tool, \eg \textsf{MongoDB}.
In order to evaluate the power consumption 
\henrique{of specific microservices}, the measurements should be filtered out to match the software application's (\eg \adapApp) containers, which are mapped to specific PIDs (Process IDs). 
\tool calculates energy consumption in the \textsf{MonitorScheduler} class
by using the equation $E = \int_{t_0}^{t_1} P(t) dt$, where $t_0$ is the timestamp at the start of the interval, $t_1$ is the timestamp at the end of the interval, and $dt$ is the 
time increment over which power is integrated~\cite{2015CengelThermodynamic}. The $P(t)$ function is calculated by using a trapezoidal rule~\cite{Burden1989} that increases the accuracy of the estimation of energy consumption over time. The choice of the rule is based on the approximation of integral $E$. The equation can be seen as follows: $E = \sum_{i=0}^{n-1} (\frac{P_i + P_{i+1}}{2} * \delta t)$,
where $P_i$ and $P_{i+1}$ are successive power measurements at times $t_i$ and $t_{i+1}$, $\delta t = t_i - t_{i+1}$ is the (constant) sampling interval, and $n$ is the number of iterations for the energy consumption monitoring (so that $n+1$ samples yield $n$ trapezoids, hence the summation index runs from $i=0$ to $i=n-1$). Practically, \tool
is configured to calculate the energy consumption by intervals of 2 seconds. After executing the \textsf{Scaphandre} tool, we found that 2 seconds is the minimum time interval that allows to collect power consumption for calculating energy consumption. 


In order to enable \adapApp to become energy-aware,
we extended its architecture 
as seen in Figure~\ref{fig:architecture} (see the grey classes inside the \adapApp component). 
%
The \textsf{EnergyConsumptionCollector} class is added and the \textsf{AdaptationScheduler} is updated to use the new collector. The \textsf{EnergyConsumptionCollector} class extends previous collectors by implementing the \textsf{IMetricsCollector} interface and by adding the  \textsf{getEnergyConsumption()} method. This method collects the energy consumed through RESTful GET requests to an energy consumption monitor, \ie  \tool, that is instrumented by a power consumption profiling tool, namely \textsf{Scaphandre}. The \texttt{max} property serves to store the maximum energy consumption obtained during the executions of the experiment; this value allows to normalise the energy consumption to make it comparable with other measurements.

The \textsf{AdaptationScheduler} class is introduced to allow the Recommender service to communicate with the energy consumption collector through the \textsf{IMetricsCollector} interface. It allows to gather energy consumption measurements when adaptations are performed. 
The analysis performed in the \textsf{AdaptationScheduler} class leverages the existing structure provided in \adapApp for the CPU load analysis, and it is extended for including the evaluation of energy consumption variations. 
Moreover, the \textsf{AdaptationScheduler} class defines conditions required to trigger adaptations. These conditions are evaluated through the \textsf{ConditionEvaluator} interface, which provides several comparison operations, including: \textsf{between two bounds, greater than, less than}, etc. The conditions are periodically checked by the \textsf{ContinuousObservationScheduler} class; this class
sets up listeners for analysing, for instance, the increase or decrease of the energy consumption at the application-level in order to ensure that adaptations are triggered as soon as the relevant conditions are met.

In order to demonstrate the validity of \tool, we conduct a \henrique{set of} controlled experiments. 
In the next section, we detail this experimentation study.


\section{Experimentation Study} \label{sec:experimentv2}
This section first presents the research questions and hypothesis for the study design. 
We then introduce the Adaptation Scenarios of the \adapApp which serve as motivating examples for validating our proposal.  
Finally, we describe the protocol we followed for conducting a controlled experiment.

\subsection{Research Questions, Hypothesis and Metrics}

The goal of the study is to validate 
the effectiveness of \tool  
to monitor energy consumption at design- and run-time. To do that, we run an experimental study in which \tool
is executed for enabling the comparison of the energy consumed by the different configurations of the \adapApp.   
From this goal, we derive the following \textbf{four research questions}:
\begin{description}
\item[RQ1:] Is the energy consumption calculated by \tool coherent \wrt other metrics (\ie CPU usage, memory consumption)?
\item[RQ2:] Does \textsf{EnCoMSAS} enable the analysis of the impact of the variants of \adapApp on its energy consumption? 
\item[RQ3:] Does \textsf{EnCoMSAS} enable the analysis of runtime adaptations of \adapApp on its energy consumption? 
\item[RQ4:] How does \textsf{EnCoMSAS} impact the energy consumption of the entire \adapApp ecosystem?
\end{description}

To answer these research questions, we conducted a controlled experiment.
We identify as an \textbf{hypothesis} that ``the service variants and adaptations at runtime influence the energy consumed by the \adapApp, and the proposed \textsf{EnCoMSAS} should enable the evaluation of this influence''. 
Moreover, the \textbf{metric} collected and analysed for this controlled experiment is focused on the energy consumption measurement (in Joules).

\subsection{Adaptation Scenarios of \adapApp}\label{sec:scenarios}

In this paper, we focus on the recommender service of  \adapApp. Thus, we first configure \adapApp for each of the four algorithms of the recommender service, and disable run-time adaptations. Then, we gather the energy consumption, and other metrics such as CPU usage and memmory consumption, of the five versions of \adapApp, \ie one version for each algorithm and one version 
with a deactivated recommender service,
and we provide statistical evidence for answering RQ1 and RQ2. 

On the other hand, in~\cite{BDGLZZ25}, different scenarios of adaptations at runtime are introduced such as unavailable resources, cyber-attack, requirements changes or traffic increase. 
To answer RQ3 and RQ4, 
we focus on the \textit{Benign Traffic Increase} scenario in which incoming traffic towards the \textsf{WebUI} increases significantly due to a genuine increase in users' interactions. In order to handle the increased load, the application is adapted to the lower-power version by switching to a more energy-efficient recommendation algorithm. Thus, we investigate the impact on the energy consumed by \adapApp under two distinct states: 
i)~\textit{without adaptation}, \ie using the default configuration (\textit{Normal} mode) of the recommender service; and ii)~\textit{with adaptation}, \ie activating the \textit{Low Power} mode to use the most energy-efficient algorithm to reduce energy consumption, or relaxing the configuration when the traffic decreases.
Moreover, iii) in order to compare what is the impact on energy consumption of \adapApp \wrt the \app (\ie the original version of \app without dynamic adaptation mechanism), we also conducted 
\henrique{an 
experiment 
adding a third mode
called \textit{Original TeaStore}}.

\subsection{Experimentation Protocol}
\label{subsec:execution}

\begin{figure}[!bhtp]
    \centering
    \includegraphics[width=1\linewidth]{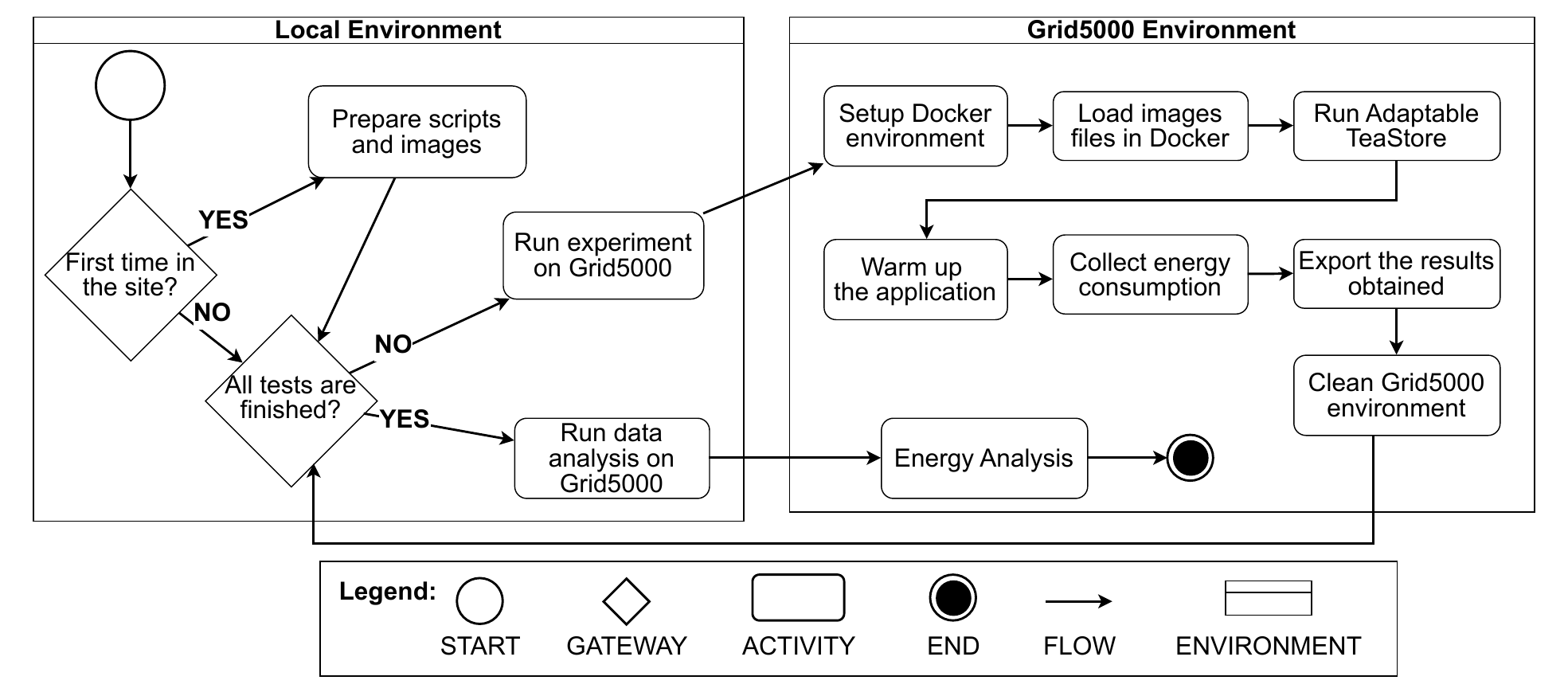}
    \caption{Step-by-step execution of the experiments}
    \label{fig:experiment}
\end{figure}

Figure~\ref{fig:experiment} illustrates the execution flow of the experiment. We considered two execution environments: a local environment, where scripts run on a local machine, and the Grid5000 environment~\cite{2005Grid}~\footnote{\url{https://www.grid5000.fr/}}.

To warm up the application, we deploy \adapApp and run a medium-intensity workload, which is characterized by an increasing workload execution for 20 seconds, with maximum peak of 1000 requests in parallel, as done in Kistowski \etal~\cite{2018TeaStore}. We then start the energy consumption monitor, that collects the energy consumed by the \adapApp during 100 seconds. Once the energy consumption data is collected, it is exported into JSON files. Finally, the Grid5000 environment is cleaned up. This process is repeated 30 times. Once the number of iterations is finished, the analysis is performed to provide evidence about the extent to which the energy consumption was affected.

We use the protocol for two distinct scenarios: without and with adaptation. \textbf{Without adaptation} (NOADAPT), the application runs without changing its recommendation algorithm. These experiments allow to analyse each algorithm individually and define the most energy-efficient one.
\textbf{With adaptation} (ADAPT), the application initially starts in the normal mode and dynamically transitions between high performance mode and the low power mode, differently from \adapApp CPU analysis that disables the high performance mode and switches between the normal mode and the low power mode.

Finally, in our extension of \adapApp, we added a flag to manage the activation of its adaptation mechanisms. It allows to ensure that the execution of the NOADAPT scenario is not affected by the presence of adaptable components, specially the analysis component from \adapApp that is permanently running. As mentioned, a third scenario is explored by using the same protocol, \ie the  \textbf{original \app}.

Regarding the data analysis, for all of the experiments, once the data are collected, we first analyze the distribution of the energy consumption (in Joules), CPU usage (in percents) and memory consumption (in Megabytes). To do that, we use the \texttt{Shapiro} statistical test, which confirmed that the measurements collected from the 30 iterations executed for the experiment presented a non-normal distribution curve. Next, we calculate the average values of the collected measurements and calculate their relative standard deviation over 30 iterations. These relative standard deviations show that the fluctuations of the values are small (i.e., less than 1), so the average values can be used as representative values for the 30 iterations. Then, we use the \texttt{Wilcoxon-Mann-Whitney} statistical test to evaluate the significant differences in the corresponding measurements of the versions of \adapApp. Finally, we use the \texttt{Pearson} statistical test in order to evaluate correlation between different measurements.

\section{Experimentation Results}\label{sec:results}
To collect the data analysed in this section, the experiments were conducted on the \textit{Nova} cluster of the Grid5000 platform (see Row 1 of Table~\ref{tab:config-clusters}). As observed, the Nova cluster is composed of 20 nodes, each of these nodes is equipped with 64 GB of RAM, 2 CPUs Intel Xeon E5-2620 with 8 cores per CPU,
598 GB of storage with 25 GB allocated per user, and a 10 Gbps network connection.
The previous configuration was used for answering the four RQs. However, for RQ2 and RQ3, we conducted the experimental study in other clusters of the Grid5000 platform (see Rows 2-5 of Table~\ref{tab:config-clusters}), it enabled analyzing how the energy consumption is affected by the physical configurations when the algorithm' variants and the dynamic adaption are examined. As observed, we used clusters with limited resources (see Rows 1 and 5 in Table~\ref{tab:config-clusters}, \eg the \textit{Econome} cluster with 8 cores/CPU, 64 GB of RAM and 2.0 TB HDD of storage) 
and clusters with more resources (see Rows 2, 3 and 4 in Table~\ref{tab:config-clusters}, \eg
the Montcalm cluster with 16 cores/CPU, 256 GiB of RAM and 960 GB SSD of storage).


\begin{table}[!b]
    \centering
    \caption{Resources of each site and cluster in Grid5000 used in the experiment}
    \label{tab:config-clusters}
    \begin{tabular}{|c|c|c|c|c|c|c|} \hline
        \textbf{Site} & \textbf{Cluster} & \textbf{Nodes} & \textbf{CPU configuration} & \textbf{Cores/CPU} & \textbf{Memory} & \textbf{Storage}  \\ \hline
        Lyon & Nova & 20 & Intel Xeon E5-2620 v4 & 8 & 64 GiB & 598 GB HDD \\ \hline
        Lille & Chifflot & 8 & Intel Xeon Gold 6126 & 12 & 192 GiB & 480 GB SSD \\ \hline
        Toulouse & Montcalm & 10 & Intel Xeon Silver 4314 & 16 & 256 GiB & 960 GB SSD \\ \hline
        Nantes & Ecotype & 48 & Intel Xeon E5-2630L v4 & 10 & 128 GiB & 400 GB SSD \\ \hline
        Nantes & Econome & 24 & Intel Xeon E5-2660 & 8 & 64 GiB & 2.0 TB HDD \\ \hline
    \end{tabular}
\end{table}


As mentioned in Section~\ref{subsec:execution}, 
we executed 30 iterations of each scenario. 
%
To simulate the user requests, we use the \textit{HttpLoadGenerator} technique~\cite{2018Joakim}.
This technique is already implemented in \adapApp and can be deployed in the Docker environment.
\adapApp proposes three workloads (\textit{High}, \textit{Medium} and \textit{Low}). 
We activate the \textit{Medium} workload to run the first scenario of the experiment and to answer RQ1 and RQ2, \ie \henrique{to} study the coherence between energy consumption \wrt the other measurements such as CPU usage and memory consumption, and study 
\henrique{to} which extent the recommender algorithms affect the energy consumption of \adapApp. Whilst for answering RQ3 and RQ4, 
\henrique{we run} the second scenario, \ie study how runtime adaptation affects energy consumption. We therefore first activate the \textit{Medium} workload and \henrique{then} increase it every second; it is \henrique{done} 
using the \textsf{Load Generator} implemented by the original \app that allows to increase quantity of activated users. Moreover, the third scenario, \ie running the orginal \app, is also executed to answer~RQ4.
Finally, the workload for each scenario was always executed on the same node to ensure the use of the same computing resources. \henrique{Doing so, allows us to keep}
the experiment controlled in order to gather energy consumption measurements that are statistically comparable.

All raw data collected in the experiments can be accessed at Zenodo (\url{https://doi.org/10.5281/zenodo.17122074}).


\subsection{RQ1: Is the energy consumption calculated by \tool coherent \wrt other metrics (\ie CPU usage, memory consumption)?}

To answer RQ1, we base our analysis on a fundamental assumption that is: ``higher resource usage generally results in increased energy demand''.  So, we
examine the relationship between the energy consumed by a software application
and its underlying hardware resource usage. 
As explained in Section~\ref{sec:experiment}, power consumption, which is used for calculating energy consumption, is primarily inferred  from the monitoring of hardware-level metrics. In particular, CPU usage represents the dominant contributor to energy consumption, whereas memory consumption typically has a limited impact with measurable effects arising primarily when memory modules are activated or deactivated~\cite{2010Tsirogiannis}. So, a correlation between energy consumption and, at least, CPU usage is expected.

Anand et al.~\cite{2023Anand} argue that energy consumption should be analyzed as an aggregate metric, independently of other performance indicators, since resource usage may depend solely on the application’s logic or its current state. However, energy consumption is also influenced by several external factors, ranging from the underlying hardware to the application itself. Therefore, it should not be interpreted as a performance- or correctness-driven metric~\cite{icse/WeberKSAS23}. Consequently, we excluded, from this analysis, throughput as a performance metric that can influence energy consumption of the \app versions.

In this section, we focus on CPU usage and memory consumption of processes. To do that, we obtain the values of these metrics for the corresponding processes by using \tool that instruments Scaphandre. Finally, Scaphandre collects and exports the values for the mentioned metrics.

\subsubsection{Analysis of CPU usage and memory consumption for the recommender algorithms of \adapApp}

\begin{figure}[!b]
    \centering
    \includegraphics[width=0.7\linewidth]{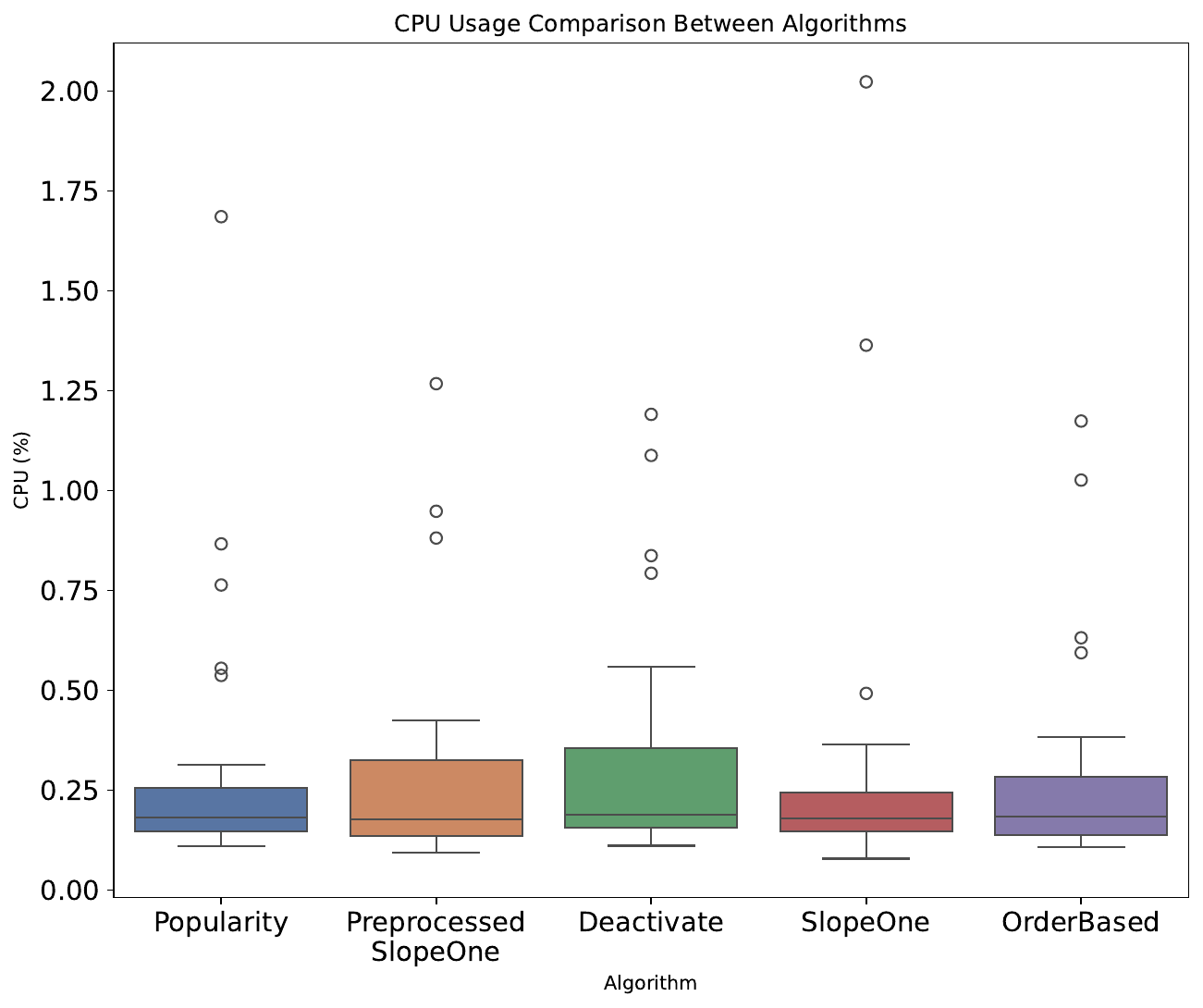}
    \caption{Distribution of the CPU usage in the Recommender service}
    \label{fig:algo-boxplot-cpu}
\end{figure}

Figure~\ref{fig:algo-boxplot-cpu} shows the distribution of the CPU usage (in percents) for the execution of the recommender algorithms of \adapApp.
Notice that the values of CPU percentage is very low, resulting from the estimation made by Scaphandre over the capacity of all the CPU cores.
The CPU usage values reveals that the \textit{Popularity} and \textit{SlopeOne} algorithms yield the lowest values of CPU usage. 
For all algorithms, the median values for CPU usage is approximately near 25\% of the data distribution.

Table~\ref{tab:cpu-usage-data-analysis} summarizes the average CPU usage per algorithm. The highest average CPU usage was observed with \textit{Deactivate} (0.323\%), followed by \textit{SlopeOne} (0.296\%), \textit{Popularity} (0.294\%), \textit{Preprocessed SlopeOne} (0.289\%), and \textit{Order-Based} (0.279\%). Regarding variability, \textit{SlopeOne} exhibits the largest standard deviation.
Notice that the CPU usage results correspond only to the Recommender microservice from \adapApp.

The Confidence Interval (CI) provides an indication of the reliability of the experimental results, representing the range within which the true value is expected to lie with approximately 95\% probability. Among the algorithms, \textit{Popularity} exhibits considerable uncertainty, as reflected by its wide confidence interval, a behavior also observed for \textit{OrderBased}. In contrast, \textit{SlopeOne} demonstrates higher estimation accuracy, evidenced by its narrower interval.

\begin{table}[ht]
    \centering
    \caption{Distribution of the CPU usage of the recommender algorithms of \adapApp. Where: \textit{AVG = Average CPU usage, CI = Confidence interval, STD = Standard deviation from the AVG values, RSEC = relative standard deviation of CPU usage}}
    \label{tab:cpu-usage-data-analysis}
    \begin{tabular}{|c|c|c|c|c|} \hline
        \textbf{Algorithm} & \textbf{AVG $\pm$ CI (95\%)} & \textbf{STD} & \textbf{RSTD}  \\ \hline
        Deactivate & 0.323$ \pm$ 0.104 & 0.279 & 0.86 \\ \hline 
        SlopeOne & 0.296 $\pm$ 0.144 & 0.386 & 1.30  \\ \hline 
        Order-Based & 0.279 $\pm$ 0.0935 & 0.251 & 0.90 \\ \hline 
        Popularity & 0.294 $\pm$ 0.116 & 0.312 & 1.06 \\ \hline 
        Preprocessed Slope One & 0.289 $\pm$ 0.0995 & 0.266 & 0.92 \\ \hline 
    \end{tabular}
\end{table}

Beyond CPU utilization, we also analyzed the memory consumption (in MB) of each algorithm within the Recommender microservice, see Table~\ref{tab:memory-usage-data-analysis}. The Deactivate algorithm exhibits the smallest variation in memory usage. However, the algorithm presents the highest average. As expected, all algorithms show an increase in consumption as the number of requests grows. Considering the confidence intervals, \textit{SlopeOne} and \textit{Popularity} emerge as the most memory-efficient, while \textit{Preprocessed SlopeOne} proves to be the most consistent. On the other hand, \textit{OrderBased} demonstrates the highest memory consumption and the least consistency, as indicated by its wide confidence interval.

\begin{table}[t]
    \centering
    \caption{Distribution of the Memory usage (in MB) of the recommender algorithms of \adapApp. Where: \textit{AVG = Average of memory usage in MB, CI = Confidence Interval, STD = Standard deviation from the AVG values, RSTD = Relative standard deviation of memory usage}}
    \label{tab:memory-usage-data-analysis}
    \begin{tabular}{|c|c|c|c|} \hline
        \textbf{Algorithm} & \textbf{AVG $\pm$ CI (95\%)} & \textbf{STD} & \textbf{RSTD}  \\ \hline
        Deactivate & 121.74 $\pm$ 10.14 & 39.93 & 0.33 \\ \hline 
        SlopeOne & 113.80 $\pm$ 25.07 & 98.71 & 0.87  \\ \hline 
        Order-Based & 118.43 $\pm$ 23.87 & 107.47 & 0.91 \\ \hline 
        Popularity & 117.01 $\pm$ 22.80 & 89.76 & 0.77 \\ \hline 
        Preprocessed Slope One & 116.45 $\pm$ 13.27  & 52.27 & 0.45 \\ \hline 
    \end{tabular}
\end{table}


To validate the relationship between energy consumption, CPU usage, and memory usage, we applied the Pearson correlation test, see Table~\ref{tab:pearson-p-values-algorithms}. The results show a strong correlation between CPU usage and energy consumption for \textit{OrderBased}, \textit{Popularity}, and \textit{Preprocessed SlopeOne}, with p-values < 0.01. In contrast, memory usage does not exhibit a significant correlation, with the lowest p-value at 0.19.

These findings indicate that CPU utilization is the primary factor influencing energy-efficiency in \adapApp, \henrique{presenting a correlation between CPU and the algorithms.}, while memory footprint has a negligible impact, \henrique{with no correlation}. 
\henrique{However, these observations cannot be generalized, as they focus specifically on \adapApp and rely on the use of Scaphandre to obtain power consumption measurements.}


\begin{table}[!bht]
\centering
\caption{Correlation between the algorithms and CPU and memory}\label{tab:pearson-p-values-algorithms}
\begin{tabular}{|c|c|c|}

\hline
 \multirow{1}{*}{} & \multicolumn{2}{c|}{\textbf{Pearson p-value}} \\

\cline{2-3}
& \textbf{\textit{CPU}} & \textbf{\textit{Memory}} \\

\hline
Deactivate & 0.067 & 0.194 \\ \hline

\cline{2-3}
SlopeOne & 0.401 & 0.446 \\ \hline

OrderBased & \textbf{0.0001} & 0.477 \\ \hline

Popularity & \textbf{0.007} & 0.400 \\ \hline

Preprocessed SlopeOne & \textbf{1e-6} & 0.658 \\

\hline
\end{tabular}
\end{table}

%

\subsubsection{Analysis of CPU usage and memory consumption for the runtime adaptations of \adapApp}

\begin{figure}[!b]
    \begin{subfigure}{0.5\textwidth}
        \includegraphics[width=0.9\linewidth,     height=6cm]{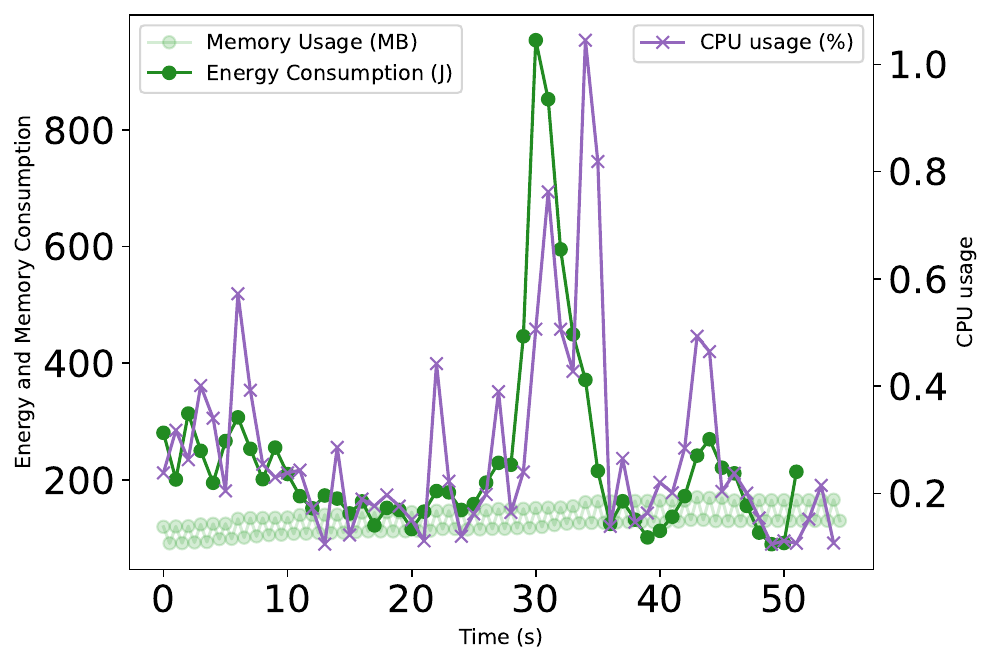}
        \caption{\textsf{ADAPT} variant}
        \label{fig:cpu-adapt}
    \end{subfigure}
    \hspace{0.4cm}
    \begin{subfigure}{0.5\textwidth}
        \includegraphics[width=0.9\linewidth,     height=6cm]{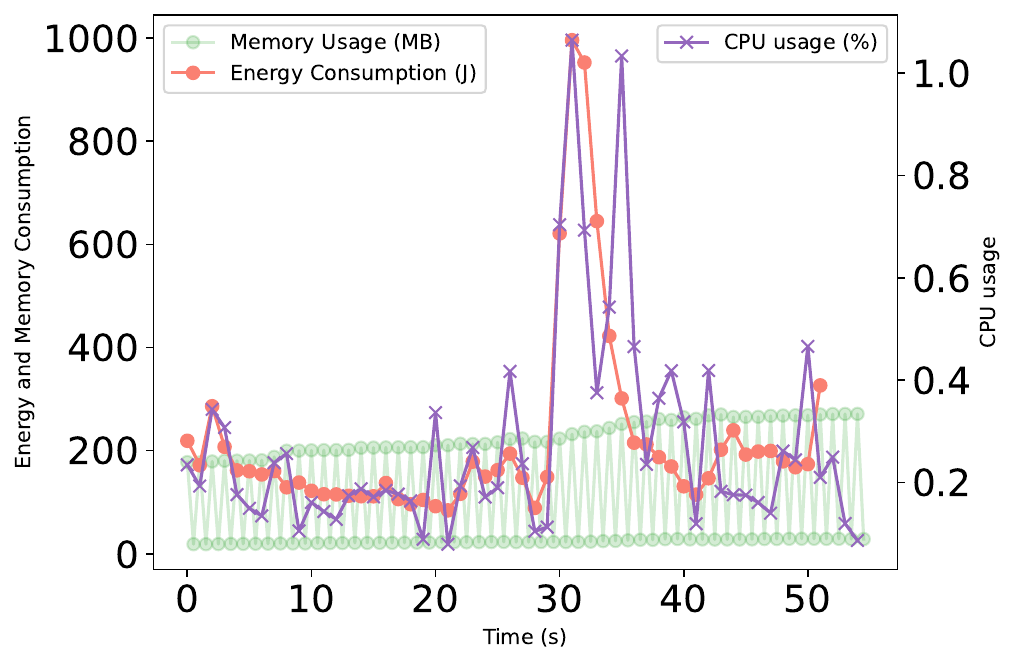}
        \caption{\textsf{NOADAPT} variant}
        \label{fig:cpu-noadapt}
    \end{subfigure}
    \caption{CPU usage and Energy Consumption of the variants}
    \label{fig:cpu-analysis}
\end{figure}

Following the analysis of the behavior of each recommendation algorithm, we conduct a comparative evaluation of the system variants with and without dynamic adaptation. The relation between CPU usage and energy consumption of the \textsf{NOADAPT} and \textsf{ADAPT} variants and their corresponding \textsf{CPU usage} is presented in Figures~\ref{fig:cpu-adapt} and~\ref{fig:cpu-noadapt}. The results present a correlation that needs to be interpreted with caution, as it is influenced by the methodology used. In our case, energy consumption was estimated using Scaphandre on the Grid5000 testbed, which bases its measurements on the virtual machine’s resource usage and the direct hardware-level power monitoring. As a result, the observed energy consumption may be partially reflective of CPU usage by design, and not an entirely independent metric. Based on the results, we conclude that energy consumption is correlated with CPU usage, as detailed in Table~\ref{fig:cpu-analysis}.

In terms of memory usage, although an overall increasing trend is observed, there is no evident correlation between memory consumption and either energy consumption or CPU usage. Figure~\ref{fig:cpu-analysis} illustrates memory consumption in the \textsf{NOADAPT} and \textsf{ADAPT} scenarios. Both configurations exhibit similar behavior, particularly within the intervals of [5-9], [27-35], and [40-47], where noticeable increases in memory usage occur. Concerning the different variants, the \textsf{NOADAPT} presents a higher standard deviation in its values, distinct from the \textsf{ADAPT} version.

\subsubsection{Correlation analysis of energy consumption with CPU usage and memory consumption for \adapApp}

Using the energy consumption data alongside memory and CPU measurements, we evaluate whether a linear correlation exists between the variants. Table~\ref{tab:pearson-p-values} presents the \textsf{p-values} obtained from the Pearson correlation test. The results indicate a strong correlation between CPU usage and energy consumption. In contrast, memory usage shows no significant correlation with energy consumption across the tested variants.

\begin{table}[!bht]
\centering
\caption{Correlation between the variants and CPU and memory}\label{tab:pearson-p-values}
\begin{tabular}{|c|c|c|}

\hline
 \multirow{1}{*}{} & \multicolumn{2}{c|}{\textbf{Pearson p-value}} \\

\cline{2-3}
& \textbf{\textit{CPU}} & \textbf{\textit{Memory}} \\

\hline
ADAPT & \textbf{5.9e-06} & 0.922\\

\cline{2-3}
NOADAPT & \textbf{4.53e-10} & 0.912 \\

\hline
\end{tabular}
\end{table}

\begin{center}
	\fbox{
		\begin{minipage}{0.9\linewidth}
\noindent \textbf{In response to RQ1:} The results indicate that CPU utilization varies across algorithms, leading to notable differences in the accuracy of the values reported by Scaphandre. Furthermore, the analysis confirms a direct association between CPU usage and energy consumption, as evidenced by Table~\ref{tab:pearson-p-values-algorithms} and~\ref{tab:pearson-p-values} and the consistent patterns observed in the Figure~\ref{fig:cpu-analysis}. With respect to memory utilization, no clear association was observed between memory consumption and energy consumption. While memory usage increased, energy consumption exhibited considerable variation, indicating that memory is not a primary determinant of the overall energy demand in this context.
		\end{minipage}
	}
\end{center}

\subsection{RQ2: Does \textsf{EnCoMSAS} enable the analysis of the impact of the variants of the \adapApp on its energy consumption?}

\begin{table}[!b]
    \centering
    \caption{Algorithm analysis. Where: $N$ = Number of distinct products, $R$ = Average number of products per user, $U$ = number of users, $I$ = Average of products per user, $C$ = Number of products in the cart, $S$ = Average of order sets per user} 
    \label{tab:algorithmcomplexity}
    \begin{tabular}{|c|c|c|} \hline
        \textbf{ALGORITHM} & \textbf{TRAINING TIME} & \textbf{EXECUTION TIME} \\ \hline
        SlopeOne & $O(UR^2)$ & $O(NR)$ \\ \hline
        OrderBased & $O(1)$ & $O(CUSIN log N)$ \\ \hline
        Popularity & $O(UI) $ & $O(NlogN)$ \\ \hline
        Preprocessed Slope One & $O(UNR)$ & $O(NlogN)$ \\ \hline
    \end{tabular}
    
\end{table}

To answer RQ2, we based the study on the hypothesis that ``\textit{the energy consumption of the variants of \adapApp depends not only on the complexity of its algorithm, but also on the computational resources}''. Thus, we first examined the source code for each recommender algorithm of the \app with the purpose of collecting information about their computational complexity. The results of this task are presented in Table~\ref{tab:algorithmcomplexity}. In this examination, we considered the training time and the execution time of each algorithm. Once the \textsf{TeaStore} application starts running, a predefined dataset is used to train each algorithm. As observed, the training time is the most demanding for the \textit{SlopeOne}, \textit{Popularity}, and \textit{Preprocessed SlopeOne} algorithms. Whilst the execution time complexity is higher in the \textit{SlopeOne} and the \textit{Order-Based} algorithms. Moreover, the \textit{Preprocessed SlopeOne} algorithm reuses the \textit{SlopeOne} algorithm during a processing step of the algorithm, without requiring the computation of the recommendation at each execution. So, the \textit{SlopeOne} is the most complex of the recommender algorithms of \adapApp.

According to the previous hypothesis, it appears that the energy consumption of recommender algorithms in different hardware configurations is independent of the algorithm complexity. 
Thus, the highest complexity algorithm, \ie \textit{SlopeOne}, would not always be the most energy demanding algorithm. In order to verify the previous claim, we monitored \adapApp after the users' requests were simulated and sent via an exposed API endpoint of the \textit{WebUI service}. The \textit{WebUI service} then calls the \textit{Recommender service} that is already configured with the algorithm that is being analyzed. For the experiment, we study each of the four algorithms of \adapApp and the version in which no recommendation is activated, \ie the \textit{deactivate} version.


As mentioned at the beginning of Section~\ref{sec:results}, we conducted the experimental study in five clusters with the different configurations of computing resources introduced in Table~\ref{tab:config-clusters}. 
Table~\ref{tab:resultRQ2} presents the distribution of the energy (see Columns 3 and 4) consumed by \adapApp in each configured cluster (see Column 1) used for executing the five versions, \ie the deactivate version and the activation of each of the four recommendation algorithms successively.

\begin{table}[!b]
\centering
\small{
\caption{The distribution of the energy consumed by the recommender algorithms, including the deactivating version. Where: \textit{DEA = Deactivate}, \textit{OB = Order-Based}, \textit{POP = Popularity}, \textit{SO = Slope One}, \textit{PSO = Preprocessed Slope One}, \emph{AEC = average of energy consumption (EC~in Joule unit}), \textit{RSEC = relative standard deviation of EC}, \textit{DEC = difference of EC vs Base}, \textit{p-value= p-value of Wilcoxon test}}\label{tab:resultRQ2}
\begin{tabular}{|c|c|c|c|c|c|c|c|c|c|c|c|c|}

\hline
 \multirow{3}{*}{\textbf{\rotatebox{90}{Cluster}}} & \multirow{3}{*}{} & 
 \multirow{3}{*}{\textbf{AEC}} & 
 \multirow{3}{*}{\textbf{RSEC}} & 
 \multicolumn{2}{c|}{\textbf{DEA}} & \multicolumn{2}{c|}{\textbf{SO}} & 
 \multicolumn{2}{c|}{\textbf{OB}} &
 \multicolumn{2}{c|}{\textbf{POP}}\\

\cline{5-12}
& &  &  & \textbf{\%DEC} & \textbf{\textit{p-value}}
& \textbf{\%DEC} & \textbf{\textit{p-value}}
& \textbf{\%DEC} & \textbf{\textit{p-value}}
& \textbf{\%DEC} & \textbf{\textit{p-value}}\\

& &  &  &  & 
&  & 
&  & 
&  & \\

\hline
\multirow{5}{*}{\textbf{\rotatebox{90}{Nova}}} 
& DEA & 18.10  & 0.64 & -- & --
& -- & --
& -- & --
& -- & --\\

\cline{2-12}
& SO & 18.51 & 0.66 & -- & 0.98
& -- & --
& -- & --
& -- & --\\

\cline{2-12}
& OB & 21.24 & 0.62 & -- & 0.18
& -- & 0.13
& -- & --
& -- & --\\

\cline{2-12}
& POP & 22.46 & 0.72 & -- & 0.38
& -- & 0.26
& -- & 0.71
& -- & --\\

\cline{2-12}
& PSO & 25.09 & 0.65 & -- & 0.14
& +36 & \textbf{0.046} 
& -- & 0.57
& -- & 0.40\\
\hline
\hline

\multirow{5}{*}{\textbf{\rotatebox{90}{Econome}}} 
& DEA & 73.9  & 0.87 & -- & --
& -- & --
& -- & --
& -- & --\\

\cline{2-12}
& SO & 67.2 & 0.67 & -- & 0.99
& -- & --
& -- & --
& -- & --\\

\cline{2-12}
& OB & 70.8 & 0.81 & -- & 0.92
& -- & 0.94
& -- & --
& -- & --\\

\cline{2-12}
& POP & 64.22 & 0.64 & -- & 0.91
& -- & 0.80
& -- & 0.81
& -- & --\\

\cline{2-12}
& PSO & 64.20 & 0.59 & -- & 0.83
& -- & 0.88
& -- & 0.94
& -- & 0.69\\

\hline
\hline

\multirow{5}{*}{\textbf{\rotatebox{90}{Ecotype}}} 
& DEA & 36.5  & 0.50 & -- & --
& -- & --
& -- & --
& -- & --\\

\cline{2-12}
& SO & 47.0 & 0.66 & +29 & \textbf{0.03}
& -- & --
& -- & --
& -- & --\\

\cline{2-12}
& OB & 35.0 & 0.58 & -- & 0.22
& -26 & \textbf{7e-3}
& -- & --
& -- & --\\

\cline{2-12}
& POP & 72.0 & 0.67 & +97 & \textbf{4e-5}
& +53 & \textbf{3e-3}
& +106 & \textbf{2e-6}
& -- & --\\

\cline{2-12}
& PSO & 37.0 & 0.56 & -- & 0.90
& -21 & \textbf{0.04}
& -- & 0.30
& -49 & \textbf{3e-5}\\

\hline
\hline

\multirow{5}{*}{\textbf{\rotatebox{90}{Montcalm}}} 
& DEA & 52.1  & 0.57 & -- & --
& -- & --
& -- & --
& -- & --\\

\cline{2-12}
& SO & 43.9 & 0.67 & -- & 0.06
& -- & --
& -- & --
& -- & --\\

\cline{2-12}
& OB & 79.9 & 0.58 & +53 & \textbf{9e-5}
& +82 & \textbf{3e-7}
& -- & --
& -- & --\\

\cline{2-12}
& POP & 37.4 & 0.47 & -28 & \textbf{6e-4}
& -- & 0.57
& -53 & \textbf{2e-9}
& -- & --\\

\cline{2-12}
& PSO & 51.5 & 0.71 & -- & 0.29
& -- & 0.37
& -36 & \textbf{3e-5}
& +38 & \textbf{0.04}\\

\hline
\hline

\multirow{5}{*}{\textbf{\rotatebox{90}{Chifflot}}} 
& DEA & 77.2  & 0.81 & -- & --
& -- & --
& -- & --
& -- & --\\

\cline{2-12}
& SO & 38.6 & 0.56 & -50 & \textbf{8e-4}
& -- & --
& -- & --
& -- & --\\

\cline{2-12}
& OB & 87.7 & 0.80 & -- & 0.40
& +127 & \textbf{1e-4}
& -- & --
& -- & --\\

\cline{2-12}
& POP & 39.3 & 0.74 & -49 & \textbf{7e-4}
& -- & 0.56
& -55 & \textbf{9e-5}
& -- & --\\

\cline{2-12}
& PSO & 50.8 & 0.71 & -- & 0.052
& -- & 0.06
& -42 & \textbf{0.01}
& +29 & \textbf{0.01}\\

\hline
\end{tabular}
}
\end{table}

We observe from Table~\ref{tab:resultRQ2} (see Columns 5 and 6) that each version of \adapApp has a different impact over different cluster configurations.
Considering that the \textit{Deactivated} version has the lowest complexity, we use it as a baseline for comparison with other algorithms (see Columns 5 and 6). Notice that \textit{Deactivate} could be the most energy-efficient in the \texttt{Nova} cluster (see Column 3 and Row 3), but it could be the most energy-demanding in the \texttt{Econome} cluster (see Column 3 and Row 8).
Moreover, we only calculated the differences (in percentage) of the energy consumption of the \adapApp versions if the \textit{p-value} obtained from executing the \texttt{Wilcoxon-Mann-Whitney} test is significant (\ie \textit{p-value < 0.05}). Thus, we observe, in the \texttt{Ecotype} cluster, that \textit{SlopeOne} and \textit{Popularity} respectively consume 29\% and 97\% more energy  than \textit{Deactivate}. However, \textit{Popularity} consumes less energy than \textit{Deactivate} for the clusters \texttt{Montcalm} and \texttt{Chifflot} saving 28\% and 49\% respectively. In addition, \textit{SlopeOne} saves 49\% in the \texttt{Chifflot} cluster.  
Finally, in the \texttt{Montcalm} cluster, \textit{OrderBase} spends 53\% more energy than \textit{Deactivate}.

Coming back to the algorithm with the highest complexity, \ie SlopeOne, we use it as baseline for comparison (see Columns 7 and 8). We observe, 
in the \texttt{Nova} cluster, that \textit{PreProcessed SlopeOne} consumes 36\% more energy than \textit{SlopeOne}.
Regarding the \texttt{Ecotype}, 
\textit{Order-Based} consumes 26\% less energy than   \textit{SlopeOne} 
and \textit{PreProcessed SlopeOne} 21\% less energy.
But, in the same cluster, \textit{Popularity} consumes 53\% more energy than the \textit{SlopeOne}.
And, in the \texttt{Montcalm} and \texttt{Chifflot} clusters,  \textit{Order-Based} consumes more energy than \textit{SlopeOne} with a difference of 82\% and 127\% respectively.

This observation points out the challenge of analysing energy consumption at process granularity.
Each cluster exhibits distinct energy consumption patterns. An analysis of the average energy consumption across scenarios shows that the \textit{Preprocessed SlopeOne} algorithm achieves the lowest average in the \texttt{Econome} cluster, the \textit{OrderBased} algorithm in the \texttt{Ecotype} cluster, the Popularity algorithm in the \texttt{Montcalm} cluster, and the \textit{SlopeOne} algorithm in the \texttt{Chifflot} cluster.

Upon executing the experiments across multiple architectures, we observed that the outcomes are highly sensitive to the chosen cluster configuration, and \tool was able to collect and detect these variations.  
Therefore, \tool can be used for performing experiments regarding energy consumption, 
however we should consider that the results must be interpreted in the context of these specific hardware configurations. It means that it is not possible to give conclusions without precising the target architecture used for the experiment. 
 

\begin{center}
	\fbox{
		\begin{minipage}{0.9\linewidth}
\noindent \textbf{In response to RQ2:} Seeing the results, we demonstrated that \tool enables the analysis of the impact on the energy consumption of the recommender service variants of \adapApp. After running the set of controlled experiments, we found that the most energy-efficient algorithm depends not only on its algorithm complexity but on the hardware configuration as well.
		\end{minipage}
	}
\end{center}

\subsection{RQ3: Does \textsf{EnCoMSAS} enable the analysis of runtime adaptations of the \adapApp on its energy consumption? }

To answer RQ3, we conduct a controlled experiment following the protocol described in Section~\ref{subsec:execution} and the scenarios introduced in Section~\ref{sec:scenarios}. Thus, the experiment is conducted under three scenarios: with adaptation (\textsf{ADAPT}), without adaptation (\textsf{NOADAPT}) and without dynamic adaptation mechanisms (\textsf{ORIGINAL TEASTORE}). 
To collect energy consumption measurements, we run \adapApp and the \app instrumented with the \tool tool.


\begin{figure}[!bp]
    \begin{subfigure}{0.5\textwidth}
        \includegraphics[width=0.85\linewidth,     height=6cm]{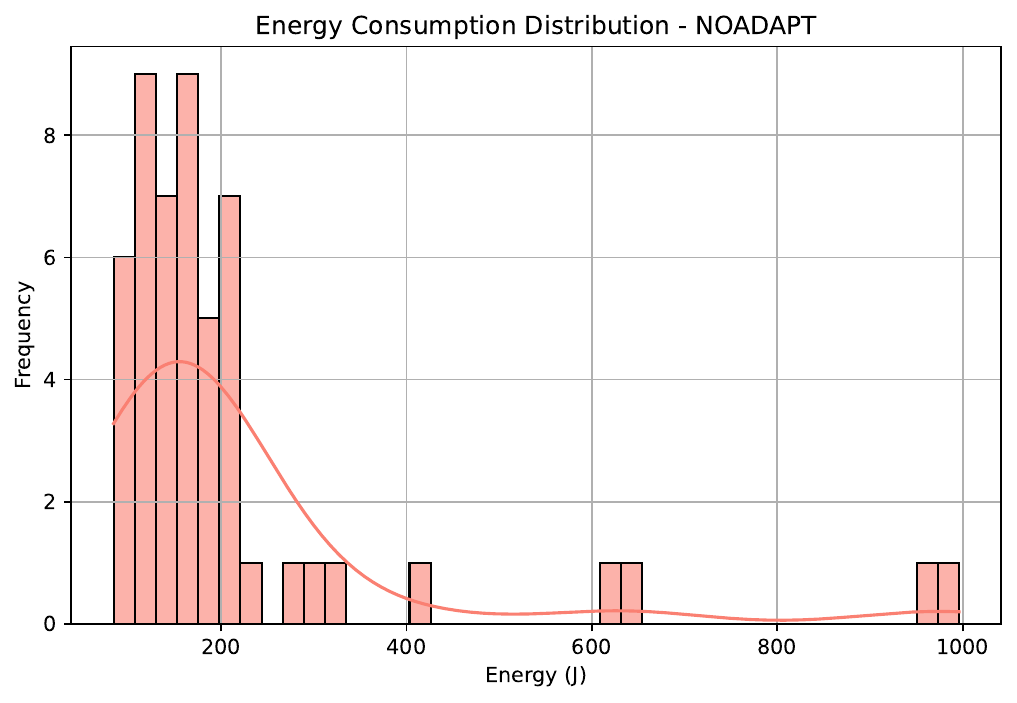}
        \caption{Energy Consumption of \textsf{NOADAPT}}
        \label{fig:dist-noadap}
    \end{subfigure}
    \begin{subfigure}{0.5\textwidth}
        \includegraphics[width=0.85\linewidth, height=6cm]{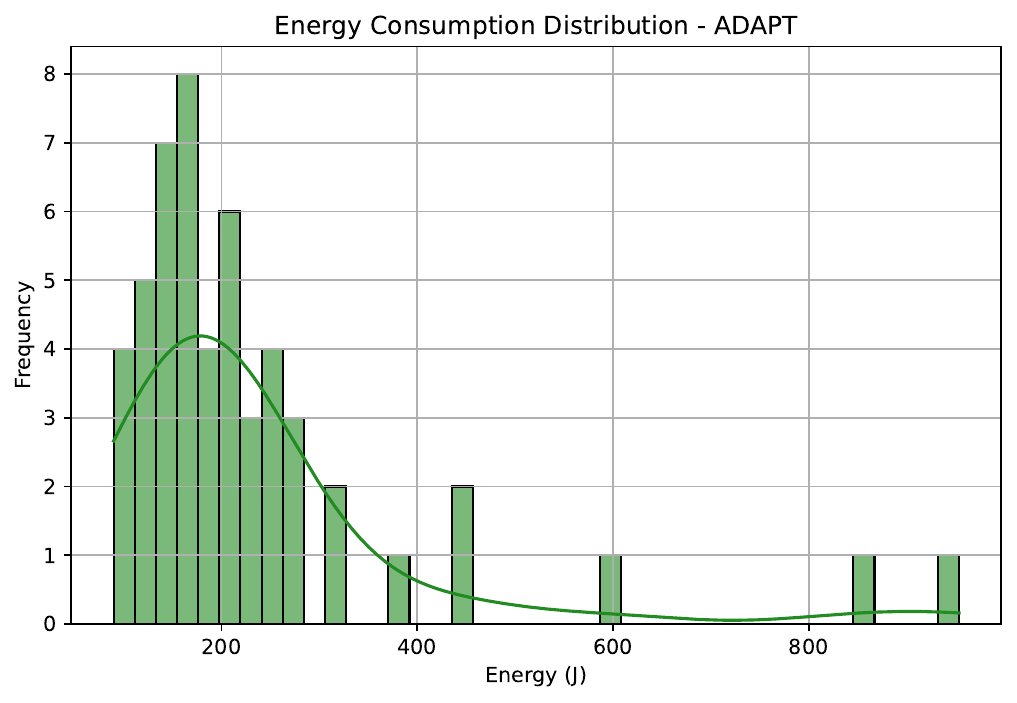} 
        \caption{Energy Consumption of \textsf{ADAPT}}
        \label{fig:dist-adap}
    \end{subfigure}
    \begin{center}
    \begin{subfigure}{0.5\textwidth}
        \includegraphics[width=0.85\linewidth, height=6cm]{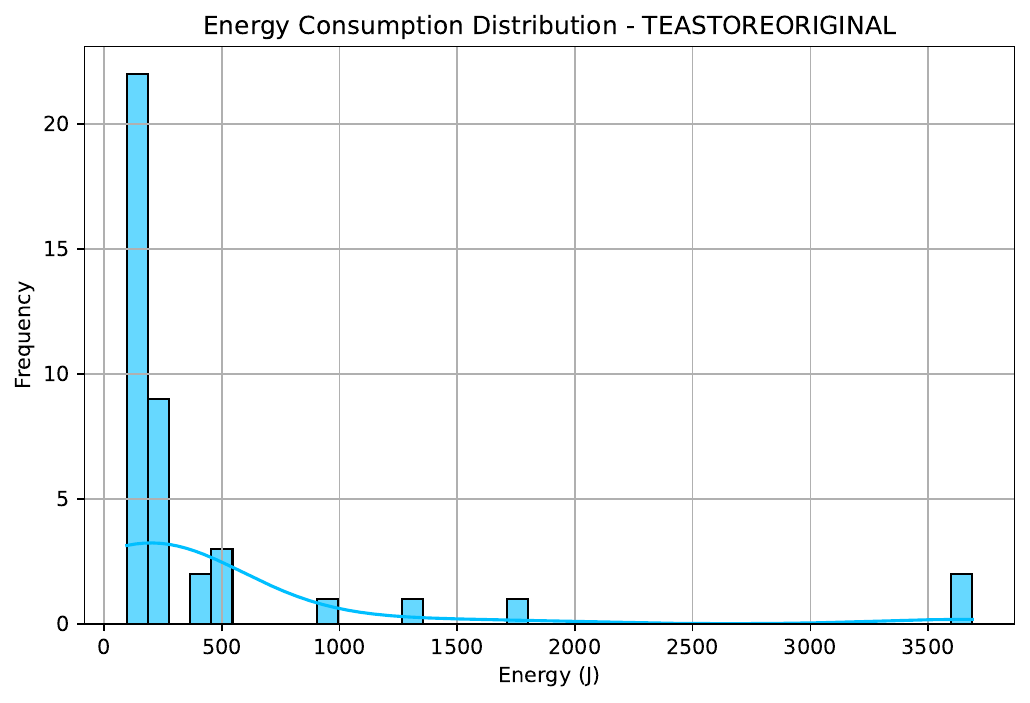} 
        \caption{Energy Consumption of \textsf{Original TeaStore}}
        \label{fig:dist-originalteastore}
    \end{subfigure}
    \caption{Distribution of the energy consumption of two scenarios}
    \label{fig:image2}
    \end{center}
\end{figure}

\begin{figure}[!bp]
    \centering
    \includegraphics[width=0.45\linewidth]{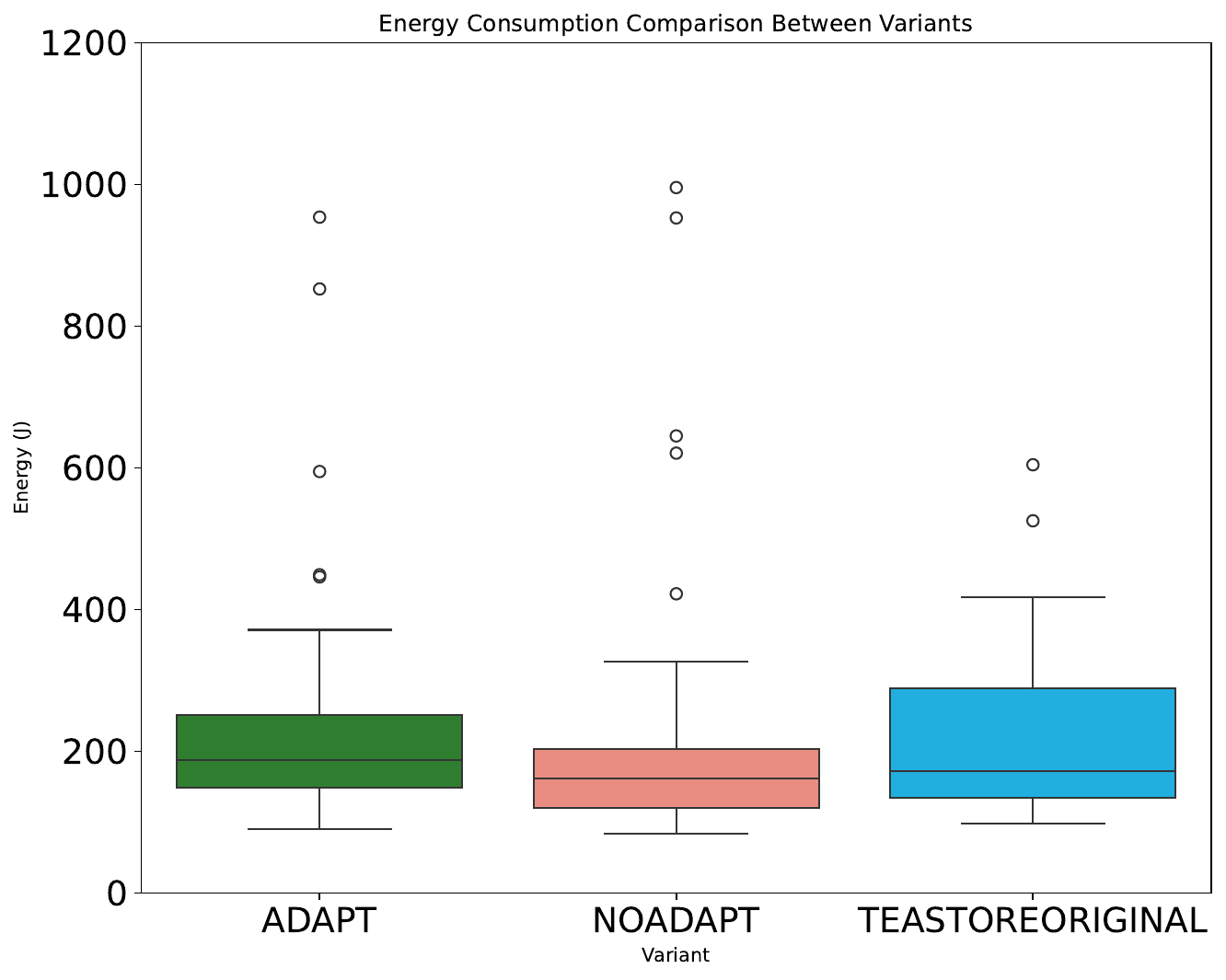}
    \caption{Data distribution of energy consumption between variants}
    \label{fig:boxplot}
\end{figure}

\begin{table}[!b]
\centering
\small{
\caption{Distribution of the energy consumed by the variants. Where: \emph{AEC = average energy consumption (EC~in Joules}), \textit{RSEC = relative standard deviation of EC}, \textit{DEC = difference of EC vs Base}, \textit{p-value= p-value of Wilcoxon test}}\label{tab:resultRQ3}
\begin{tabular}{|c|c|c|c|c|c|c|c|c|}

\hline
 \multirow{3}{*}{\textbf{\rotatebox{90}{Cluster}}} & \multirow{3}{*}{} & 
 \multirow{3}{*}{\textbf{AEC}} & 
 \multirow{3}{*}{\textbf{RSEC}} & 
 \multicolumn{2}{c|}{\textbf{ORIGINAL TEASTORE}} & \multicolumn{2}{c|}{\textbf{ADAPT}} \\

\cline{5-8}
& &  &  & \textbf{\%DEC} & \textbf{\textit{p-value}}
& \textbf{\%DEC} & \textbf{\textit{p-value}}\\

& & & & & & & \\

\hline
\multirow{3}{*}{\textbf{\rotatebox{90}{{\scriptsize Nova}}}} 
& ORIGINAL TEASTORE & 454.35  & 1.78 & -- & --
& -- & --\\

\cline{2-8}
& ADAPT & 233.38 & 0.71 & -- & 0.838
& -- & --\\

\cline{2-8}
& NOADAPT & 216.85 & 0.87 & -- & 0.277
& -- & 0.074\\
\hline

\hline
\multirow{3}{*}{\textbf{\rotatebox{90}{{\scriptsize Econome}}}} 
& ORIGINAL TEASTORE & 37.57  & 0.46 & -- & --
& -- & --\\

\cline{2-8}
& ADAPT & 75.27 & 0.60 & +100 & \textbf{3e-8}
& -- & --\\

\cline{2-8}
& NOADAPT & 69.32 & 0.66 & +85 & \textbf{5e-5}
& -- & 0.355\\
\hline

\hline
\multirow{3}{*}{\textbf{\rotatebox{90}{{\scriptsize Ecotype}}}} 
& ORIGINAL TEASTORE & 28.19  & 0.57 & -- & --
& -- & --\\

\cline{2-8}
& ADAPT & 38.08 & 0.89 & -- & 0.67
& -- & --\\

\cline{2-8}
& NOADAPT & 62.13 & 0.78 & +120 & \textbf{7e-5}
& +63 & \textbf{0.0001} \\
\hline

\hline
\multirow{3}{*}{\textbf{\rotatebox{90}{{\scriptsize Montcalm}}}} 
& ORIGINAL TEASTORE & 31.32  & 0.55 & -- & --
& -- & --\\

\cline{2-8}
& ADAPT & 56.29 & 0.48 & +80 & \textbf{1e-6}
& -- & --\\

\cline{2-8}
& NOADAPT & 45.87 & 0.50 & +46 & \textbf{0.001}
& -18 & \textbf{0.009}\\
\hline

\hline
\multirow{3}{*}{\textbf{\rotatebox{90}{{\scriptsize Chifflot}}}} 
& ORIGINAL TEASTORE & 33.92  & 0.92 & -- & --
& -- & --\\

\cline{2-8}
& ADAPT & 71.46 & 0.71 & +111 & \textbf{7e-6}
& -- & --\\

\cline{2-8}
& NOADAPT & 31.33 & 0.49 & -- & 0.55
& -56 & \textbf{5e-8}\\
\hline

\hline
\end{tabular}
}
\end{table}





According to the feature model of \adapApp introduced in Figure~\ref{fig:fm}, the recommender service can dynamically adapt itself by activating/deactivating only two available modes, \ie \textit{Normal} and \textit{High Performance} modes that respectively execute the \textit{Slope One} and the \textit{Preprocessed Slope One} algorithms. 
Regarding the \textsf{NOADAPT} scenario, we chose the default configuration of \adapApp which is the \textit{Normal Mode}. It means that the \textsf{NOADAPT} scenario runs the \textit{Slope One} algorithm and no adaptations are executed.
Whilst for the \textsf{ADAPT} scenario, we configured \adapApp to use the \textit{Slope One} algorithm for the \textit{Normal Mode}.
For the \textit{High Performance Mode}, we retained the \textit{Preprocessed Slope One} algorithm, as it is already available for dynamic activation/deactivation. The other algorithms are not activated for dynamic adaptation.
The microservice switches between these two modes according to the energy consumption of the service. The first adaptation is triggered when the current energy consumed by \adapApp is 50\% more than the previous energy consumption collected every 2 seconds. The second adaptation is triggered when the energy consumption is below 50\%. It is worth to mention that these two adaptation rules are introduced only for illustration purpose, a deeper analysis should be performed to determine practical adaptation rules.


As observed in Figure~\ref{fig:image2}, \textsf{ADAPT} tends to consume more energy than \textsf{NOADAPT} and the \textsf{Original TeaStore}, not considering the outlier values.
As mentioned, at runtime, the recommender service dynamically selects different algorithms based on the energy consumption,
\ie reducing the energy consumption in response to an increasing workload and relaxing the energy consumption restriction when the workload is reduced.
So, the difference in the energy consumption distribution is attributed to the adaptive behaviour of the recommender service in the \textsf{ADAPT} scenario of the application.
On the other hand, the energy consumption distribution of the \textsf{Original TeaStore} shows lower overall consumption compared to both the \textsf{ADAPT} and \textsf{NOADAPT} variants. Notably, it also exhibits a higher concentration of outliers at elevated values, over the interval of [500J, 2000J], see Figure~\ref{fig:dist-originalteastore}. 
The dynamic adaptation behaviour of the \textsf{ADAPT} version is particularly evident in the energy interval between [300J-600J], as shown in the Figures~\ref{fig:dist-noadap} and~\ref{fig:dist-adap}. In this range, the adapted version consistently demonstrates higher energy consumption than the other versions, whereas the non-adaptive version shows a gradual decrease in energy consumption.

Figure~\ref{fig:boxplot} and Table~\ref{tab:resultRQ3} 
present the distribution of energy consumption across these three scenarios in different clusters.
We observe from the \texttt{Nova} cluster that the lowest recorded energy consumption occurs in the non-adaptive scenario, while the highest consumption appears in the \textsf{Original TeaStore}, followed by the adaptive scenario. However, no significant differences were calculated between the execution of these three versions (see Column 6 and Rows 4 and 5),  we cannot then conclude whether a version is more, or less, energy-efficient.
In all other clusters, the \textsf{Original TeaStore} consistently demonstrated the highest energy efficiency, with the exception of the \texttt{Chifflot} cluster, where it ranked as the second most energy-efficient version. 
However, as we can observe (see Column 6 and Row 17), for the \texttt{Chifflot} cluster, a significant difference between the \textsf{Original TeaStore} and \textsf{NOADAPT} was not obtained, so we cannot conclude that the \textsf{Original TeaStore} is more energy efficient than \textsf{NOADAPT}, and vice-versa.
Moreover, we can observe from Columns 7 and 8 that \textsf{NOADAPT} consumes 63\% more energy than \textsf{ADAPT} for the \texttt{Ecotype} cluster, however for the \texttt{Montcalm} and \texttt{Chifflot} clusters,  \textsf{NOADAPT} is more energy efficient than \textsf{ADAPT} by saving 18\% and 56\% respectively. \henrique{All the previously reported differences were obtained as significant, considering the \textsf{p-value} < 0.05.}
As observed in RQ2, upon executing the experiments across multiple architectures, we observed that the outcomes are sensitive to the chosen cluster configuration, in particular for the \texttt{Ecotype} cluster. The \tool tool was able to collect and detect these variations.

\begin{center}
	\fbox{
		\begin{minipage}{0.9\linewidth}
\noindent \textbf{In response to RQ3:} Seeing the results, we demonstrated that \tool enables the analysis of energy consumption for the run-time adaptations of \adapApp. After running the controlled experiment, we found, as in RQ2, that the most energy-efficient version depends on the hardware configuration. However, we discovered that
there is a tendency, 4 out of 5 clusters, in which the \textsf{Original TeaStore} consumes less energy than the \adapApp for the \textsf{NOADAPT} and \textsf{ADAPT} versions.
		\end{minipage}
	}
\end{center}

\subsection{RQ4: How does \textsf{EnCoMSAS} impact the energy consumption of the entire \adapApp ecosystem?}

To answer RQ4, we conduct an experimental study for analyzing  the energy consumed by all the components involved in the \adapApp architecture.
Thus, the study involved four components:  
\textsf{EnCoMSAS}, \textsf{Scaphandre}, \textsf{MongoDB} and 
the recommender service configured in \adapApp for the \textsf{ADAPT} version. 

\begin{figure}[!b]
    \centering
    \includegraphics[width=0.45\linewidth]{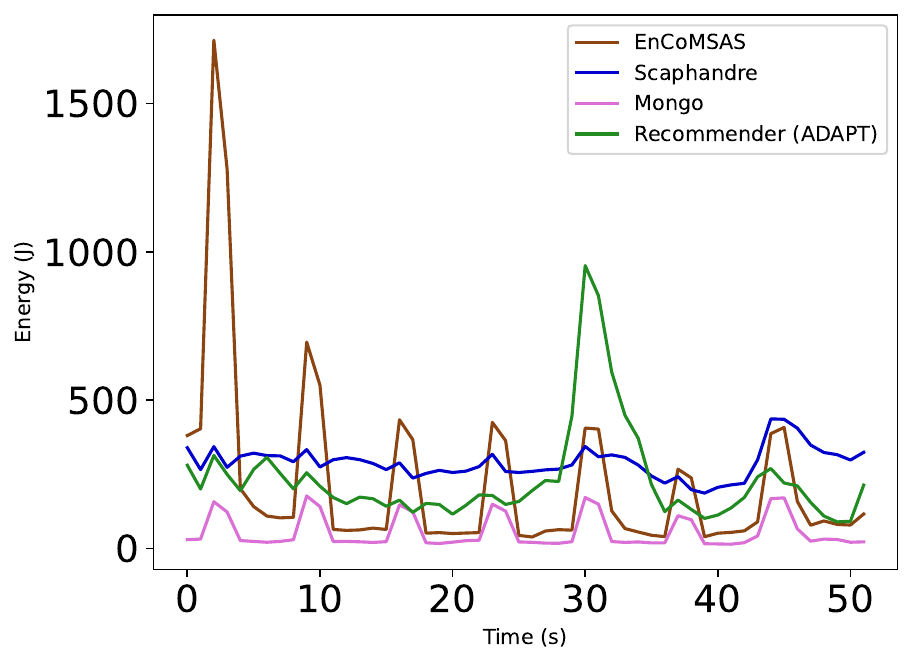}
    \caption{Energy Consumption of the components of the \adapApp architecture}
    \label{fig:energyconsumptioncomponents}
\end{figure}

Figure~\ref{fig:energyconsumptioncomponents} presents a comparison of energy consumption between these components.
%
Notably, \textsf{EnCoMSAS} exhibits periodic spikes in energy consumption, which align with query operations (read and write) to \textsf{MongoDB}. This pattern is consistently observed across both components, indicating that database access is a primary contributor to energy variations.
Moreover, \textsf{Scaphandre} shows the highest overall energy consumption among the components. This is expected, as it is responsible for performing energy consumption estimation, a task that involves continuous monitoring and processing of system metrics.
It is important to notice that \textsf{Scaphandre} measures the energy consumption of all processes running on the virtual machine. Therefore, when additional services, such as the \textsf{WebUI} are included in the analysis, there is no incremental energy overhead from \textsf{Scaphandre}, as the monitoring is already active. Similarly, \textsf{EnCoMSAS} only requires updates to its configuration to begin analyzing new processes or microservices without incurring significant additional energy costs.

Table~\ref{tab:comparative-microservices} presents the energy consumption of each microservice, including the \tool components. As observed, the \textit{WebUI microservice} emerges as the most energy-demanding component of the architecture, followed closely by the \textit{Image Provider}, which is responsible for rendering the user interface and system images, respectively. Regarding the \tool stack itself, comprising \textsf{Scaphandre} and \textsf{MongoDB}, the total energy consumption is approximately 558 J. This positions \tool as the second most energy-demanding microservice in the system, surpassed only by the \textit{WebUI TeaStore} microservice.

\begin{table}[t]
    \centering
    \caption{Overall energy consumption of the microservices. Where: \emph{AEC = average energy consumption (EC~in Joules}), \textit{RSEC = relative standard deviation of EC}}
    \label{tab:comparative-microservices}
    \begin{tabular}{|c|c|c|c|c|c|} \hline
        \textbf{Service} & \textbf{AVG} & \textbf{RSEC} & \textbf{Min (J)} & \textbf{Max (J)} \\ \hline
        WebUI & 1113.42 & 0.90 & 123.84 & 3118.82 \\ \hline
        Image Provider & 522.37 & 1.26 & 80.91 & 2197.17 \\ \hline
        Persistence & 441.94 & 0.88 & 18.67 & 1204.59 \\ \hline
        Auth & 414.18 & 1.38 & 70.35 & 2235.4 \\ \hline
        Recommender (NOADAPT) & 216.84 & 0.87 & 84.17 & 995.68  \\ \hline
        Registry & 116.66 & 1.26 & 16.81 & 616.56 \\ \hline
        \hline
        \tool & 215.26 & 1.47 & 32.73 & 1756.1 \\ \hline
        Scaphandre & 286.15 & 0.18 & 194.21 & 436.72 \\ \hline
        MongoDB & 56.66 & 1.01 & 14.87 & 180.05 \\ \hline
    \end{tabular}
\end{table}

Examining the differences across algorithms (see Table~\ref{tab:correlaction-encomsas}), all \adapApp microservices show significant variation when compared to the \tool components. Scaphandre is an exception, exhibiting no significant difference relative to the WebUI or Persistence microservices, below 0.05. Among the microservices, WebUI demonstrates the largest deviation (consuming more than +417\%, over Monitor, and +1865\%, over MongoDB), followed by the Image Provider consuming more than +822\% than the MongoDB.

\begin{table}[!t]
\centering
\caption{Difference between each microservice. Where: \textit{DEC = difference of EC vs Base}, \textit{p-value= p-value of Wilcoxon test}}\label{tab:correlaction-encomsas}
\begin{tabular}{|c|c|c|c|c|c|c|}

\hline
 \multirow{1}{*}{} & \multicolumn{2}{c|}{\textbf{\tool}} & \multicolumn{2}{c|}{\textbf{Scaphandre}} & \multicolumn{2}{c|}{\textbf{MongoDB}} \\

\cline{2-7}
& \textbf{\textit{\%DEC}} & \textbf{\textit{p-value}}
& \textbf{\textit{\%DEC}} & \textbf{\textit{p-value}}
& \textbf{\textit{\%DEC}} & \textbf{\textit{p-value}}\\

\hline
WebUI & +417.24 &  \textbf{2e-9}
&  &  0.072
& +1865 &  \textbf{1e-16} \\ \hline

\cline{2-3}
Image Provider & +143 &  \textbf{2e-5}
& +83 &  \textbf{0.044}
& +822 &  \textbf{6e-13} \\ \hline

Persistence & +105 & \textbf{0.0026}
&  &  0.25
& +680 &  \textbf{1e-9} \\ \hline

Auth & +92 & \textbf{0.0005}
& +45 &  \textbf{0.0006}
& +631 & \textbf{7e-12} \\ \hline

Recommender (NOADAPT) & +0.73 & \textbf{0.0007} 
& -24 &  \textbf{5e-8}
& +282.70 &  \textbf{3e-14} \\ \hline

Registry & -46 & \textbf{1e-5}
& -59 &  \textbf{5e-10}
& +106 &  \textbf{0.0009} \\

\hline
\end{tabular}
\end{table}

\begin{center}
	\fbox{
		\begin{minipage}{0.9\linewidth}
\noindent \textbf{In response to RQ4:} The results show that the energy consumption introduced by \tool is comparatively modest when considered across the entire set of \app microservices. However, when \tool is used to monitor a single microservice in isolation, its overhead may outweigh the energy consumption of the target service itself. In such cases, the deployment of \tool may not be beneficial, as the monitoring cost could exceed the energy savings achieved.
		\end{minipage}
	}
\end{center}


\section{Threats to Validity}\label{sec:threats}

In this section we discuss potential threats to the validity of our experiment by grouping them into construct, internal, conclusion, and external validity threats~\cite{wohlin2012experimentation}.

\noindent \textbf{Internal validity:} 
%
To collect energy consumption measurements, we employed \tool, a tool validated in our experiments. Since the measurement process itself could potentially introduce bias, the design and implementation of \tool were carried out independently from the experimental setup. This separation ensures that the measurement framework does not interfere with or bias the results reported in this study.

\noindent \textbf{Construct validity:} 
%
For the design of the experiment, both the client and server were deployed on the same node. This setup inherently constrains the evaluation, as client requests directly influence the performance of the server. However, this threat is partially mitigated by \tool, which enables fine-grained measurement of energy consumption at the container level, thereby isolating the consumption of the containers relevant to the experiment.

To enhance the robustness of our findings, we plan to extend future experiments to different cloud applications and domains, thereby assessing the generalizability of our approach. Another potential threat arises from fluctuations in energy measurements, which may bias results and conclusions. To address this issue, each experiment was executed 30 times, and average values were computed to smooth out variability. Additionally, application performance may be affected by residual load from previous requests. To mitigate this threat, energy measurements were collected across different nodes under equivalent conditions.



\noindent \textbf{External validity:} 
%
Since our experiments were conducted on specific hardware configurations and workloads, the results cannot be directly generalized to other environments. This threat is partially mitigated by employing diverse hardware resources and leveraging the range of workloads provided by \app, which increases the representativeness of the evaluation.

\section{Related Work}\label{sec:relatedwork}
The related work 
can be broadly categorised into two areas: research focused on designing energy-efficient architectures that offer tactics and patterns for building greener Cloud applications, and research exploring runtime adaptations that consider energy-consumption metrics. The latter remains less explored, especially when considering performance-related metrics such as energy consumption rather than CPU load or memory usage.

\paragraph{Energy-efficient architectures.}

The energy consumption of Cloud applications has been widely explored across various fields in the literature. Some studies estimate energy usage based on the application’s consumption of machine resources such as CPU and memory. Others take a broader approach by measuring the host machine's overall energy consumption to infer the energy footprint of the application.

In~\cite{2025Xiao}, Xiao~\etal analyse tactics and patterns from the literature related to energy consumption in microservices, selecting three tactics and three patterns, and subsequently examining the trade-offs with maintainability and performance. In~\cite{2025Noureddine}, Noureddine and Le~Goaer follow the same principle, analysing the  impact of design patterns on energy consumption. In contrast, our work presents an in-depth analysis of adaptation at the functional level of a microservice.

In~\cite{2025Werner}, Werner~\etal propose a Kubernetes-based framework for evaluating Cloud applications using sustainability, quality, and performance metrics across different architectural layers. Their analysis offers a high-level view of application metrics. In contrast, our work provides a more fine-grained, function-level analysis focused solely on energy consumption. We extended a Cloud application that supports adaptations based on CPU usage to also support adaptations triggered by energy consumption at runtime, offering deeper insight beyond the broader design-time architectural analysis presented in their study.

\paragraph{Run-time Application Adaptations.}

Existing self-adaptive systems that address energy consumption typically base decisions on host-level resource usage. For instance, in~\cite{2019Malik}, Malik~\etal aim to reduce energy consumption by scaling resources based on performance metrics and power meter readings. Berkane~\etal propose in~\cite{2022Berkane} some elastic scaling rules informed by user activity and server utilisation, while Huber~\etal ~\cite{2011Huber} focus on minimising resource usage while preserving response time.

Our work differs by targeting energy consumption at the application level, independently of infrastructure metrics. This approach allows us to isolate and understand the impact of runtime adaptations driven solely by energy usage, particularly through the selection of algorithms designed to optimise energy efficiency of one microservice.

\section{Conclusion}\label{sec:conclusion}
Understanding the energy consumption of applications is crucial for developing energy-efficient solutions. 
Through the proposed analysis, developers can make better decisions 
about which architectural adaptations best suit specific scenarios. For instance, in an online store environment, like in this work, periods of high demand such as peak sales seasons may call for performance-optimised configurations, even at the cost of increased energy usage. By balancing performance requirements with energy-consumption considerations, it becomes possible to create adaptable applications that are both energy efficient and responsive to changing conditions.
Therefore, in this paper, we introduced the \textsf{EnCoMSAS} tool that allows to monitor the energy consumed by software applications at runtime, and to trigger dynamic adaptation when needed. Moreover, \textsf{EnCoMSAS} is able to set up the monitoring frequency.

Based on the experiments we conducted to evaluate the effectiveness of \textsf{EnCoMSAS}, we first assess 
the extent to which
the algorithms implemented by \adapApp for its recommender service and the dynamic adaptation we propose
affect the CPU usage and memory consumption.
From these experiments, we also analyse the energy consumption of each algorithm in different cluster configurations.
We found that the most energy efficient algorithm is  \textsf{Slope One} for a specific cluster configuration.
A \henrique{set of} controlled experiment was also performed to evaluate the extent to which the adaptative version of \adapApp affects its energy consumption with respect to its non adaptative version and to the \app.
It shows the effectiveness of \textsf{EnCoMSAS} for running such experiments in which the 
\henrique{energy consumption}
is required.
Lastly, we 
\henrique{compare}
the energy consumption of \textsf{EnCoMSAS} 
with the entire \adapApp energy consumption. However, more experiments are needed in order to understand the impact of \textsf{EnCoMSAS} on the quality of service attributes of software applications, such as performance and energy efficiency itself, once we prove that the energy consumption does not depend only on the algorithms, but also on the configuration of the running hardware.

Another future work is to study what is the impact of having an external component, \ie not part of the \adapApp architecture,  for the adaptation analysis and planning steps,
as this is made in 
the MAPE-K loop~\cite{mapek}. Centralising the adaptation logic in a single internal service, as observed in our experiment, 
may limit the effectiveness of the energy efficiency. By decoupling adaptation decisions from the core services, it becomes possible to make more informed and flexible energy-aware adjustments across the application ecosystem. 


\section*{Acknowledgements}
The authors thank Yakine Klabi and Sara Kossentini for their work on the study of the variants of the recommendation service for \adapApp \henrique{(it was made during their Master research project at Télécom SudParis)}, and Thierry Zheng for reproducing the experiments from this paper \henrique{under} different types of architectures in Grid5000.
This research was produced within the framework of Energy4Climate Interdisciplinary Center (E4C) of IP Paris and Ecole des Ponts ParisTech. This research was supported by 3rd Programme d'Investissements d'Avenir [ANR-18-EUR-0006-02]. 
This work received funding from the France 2030 program, managed by the French National Research Agency under grant agreement No. ANR-23-PECL-0003.

\nocite{*}

\bibliographystyle{eptcs}
\bibliography{references}

\end{document}